%% file: fermion.tex
\def\CC{{\rm\kern.24em \vrule width.04em height1.46ex depth-.07ex
\kern-.30em C}}
\def\P{{\rm I\kern-.25em P}}

\documentstyle[aps,pra,multicol,epsf]{revtex}

\begin{document}
\draft
\title
{Quantum Entanglement  in Fermionic Lattices 
 }
\author{Paolo Zanardi}
\address{ Institute for Scientific Interchange  Foundation, 
Viale Settimio Severo 65, I-10133 Torino, Italy\\
 and  Unit\`a INFM, Politecnico di Torino,
Corso Duca degli Abruzzi 24, I-10129 Torino, Italy
}
\maketitle
 \begin{abstract}
The Fock space of a system of indistinguishable particles  is isomorphic
(in a non-unique way) to the state-space of a composite i.e., many-modes, quantum system.
One can then discuss quantum  entanglement
for fermionic as well as bosonic systems.
We exemplify the use of this  notion -central in quantum information -  by studying some e.g., Hubbard,
lattice fermionic  models relevant to condensed matter physics.

\end{abstract}

\begin{multicols}{2}
\narrowtext
A big deal of attention have been recently devoted to the notion of quantum entanglement
\cite{ent}. As a matter of fact this  fashion is
mostly due to the vital role that such notion is generally believed
to play in Quantum Information Processing (QIP) protocols \cite{QC}.
The  definition of entanglement
relies on the tensor-product structure of the state-space
of a composite quantum system \cite{peres}.

On the other hand such a tensor product structure is not present
in a large class of systems of major physical interest: ensembles 
of {\em indistinguishable} particles.
Indeed in this case it is known -- basically since
the birth of quantum theory-- that the state-space
associated with $N$ subsystems is constrained to be {\em subspace} of the
$N$-fold tensor product. Depending on the bosonic or fermionic 
nature  of the subsystems   one has to select the totally symmetric or  antisymmetric subspace.
This is an instance  of  a {\em superselection} rule \cite{selection} 
i.e., a fundamental limitation on the possibility of preparing a given state.

We see therefore that  the existence of  quantum {\em statistics} \cite{peres}
makes the notion of entanglement problematic for systems of made of indistinguishable
subsystems e.g., particles. 
Notice that, even  if one allows for 
general {\em parastatistics} \cite{peres}, most of the product states $\otimes_{i=1}^N |i\rangle$ do {\em not } belong
to the physical state-space at all in that they do not have a proper
transformation  under permutations of ${\cal S}_N$ i.e., they do not belong to an ${\cal S}_N$-irrep \cite{ex}.

Very recently  some authors addressed the issue of entanglement
(or more generally quantum correlations) in system of two fermions \cite{john}
and bosons \cite{li}. Their approach  appears to be the natural
generalization of the one  used for the distinguishable particles.

In this paper we shall 
tackle the problem of the    relation between entanglement and quantum statistics
from a rather different perspective
 based on  entanglement relativity as discussed in Ref. \cite{virtual}
We shall mostly
focus on the fermionic case \cite{kit}.
In particular we shall analyze the local i.e., on-site, entanglement
associated with simple fermionic models on a lattice.

{\em Fermions and Qubits.}
Let us start be recalling basic kinematical facts about many fermion systems.
Let $h_L:=\mbox{span}\{|\psi_l\rangle\}_{l\in{\bf{N}}_L}$ (${\bf{N}}_L:=\{1,\ldots,L\}$) 
be an $L$-dimensional {\em single particle}
state space.
The labels in ${\bf{N}}_L$ will be referred to as {\em sites} and the associated single-particle
wavefunctions will be thought of as describing a (spatially) localized state.
Accordingly the set of the $l$'s will be referred to as the {\em lattice}.

Let  ${\cal P}_L$ (${\cal P}_L^N$) denote  the whole (with $N$ elements) family of subsets  of ${\bf{N}}_L.$
For any $A:=\{j_1,\ldots,j_N\}\in{\cal P}_L^N$
we define the anti-symmetrized state-vector
\begin{equation}
|A\rangle:= \frac{1}{\sqrt{N!}} \sum_{P\in{\cal S}_N} (-1)^{|P|} \otimes_{l=1}^N |\psi_{j_{P(l)} }\rangle.
\label{basis}
\end{equation}
The $|A\rangle$'s are an orthonormal set.
The state-space ${\cal H}_L(N)$ associated with $N$ ({\em spinless}) fermions with 
single-particle wavefunctions belonging to $h_L$
is given by the totally anti-symmetric subspace of $h_L^{\otimes\,N}$ i.e., 
$H_L(N):=
\mbox{span}\{|A\rangle\,/\, A\in {\cal P}_L^N\}.$ 
The fermion number ranges from $0$ to $L,$ the total {\em Fock} space is obtained as a direct sum
of the fixed number subspaces i.e.,
${\cal H}_L=\oplus_{N=0}^L H_L(N)=\mbox{span}\{|A\rangle\,/\, A\in {\cal P}_L\}.$
From the well-known relation  
%\begin{equation}
$\mbox{dim} {\cal H}_L=\sum_{N=0}^L \mbox{dim}  H_L(N)= \sum_{N=0}^L \pmatrix{ L\cr N} =2^L,
$%\end{equation}
it follows that the fermionic Fock  space  is 
isomorphic to a $L$-qubits space, each qubit being associated 
with a site \cite{bose}. The latter isomorphism is realized by the the mapping 
\begin{equation}
\Lambda\colon {\cal H}_L\rightarrow (\CC^2)^{\otimes\,L}\colon 
|A\rangle \rightarrow \otimes_{l=1}^L |\chi_A(l)\rangle,
\label{mapping}
\end{equation}
where $\chi_A\colon{\bf{N}}_L\rightarrow \{0,\,1\}$ is the {\em characteristic } function of $A.$
Clearly $\Lambda(|A\rangle)$ is nothing but a $N$-qubit basis state having in the $j$-th site
a one (zero) if $j\in A$ ($j\notin A$). 
The $0$-particle state $|\emptyset\rangle$ 
is mapped by $\Lambda$ 
onto $|0\rangle:= |0\rangle^{\otimes\,L};$ thus latter vector is referred to as the {\em vacuum}.

In our considerations, once ${\cal H}_L$ is endowed by  $\Lambda$
with a  multi-partite structure,   tensor
products of individual single-particle spaces are 
not relevant anymore.  
To exemplify this point let us consider the case $L=3.$
It is not difficult to see  that all the states in
${\cal H}_3(2),$ seen as elements of $h_3^{\otimes\,2},$  have the {\em same} entanglement.
Indeed all of them can be written as $|a\rangle\otimes|b\rangle- |b\rangle\otimes|a\rangle,$
for suitable $|a\rangle$ and $|b\rangle $ \cite{zap}.
On the other hand both the  "separable" state $|1\rangle\otimes|1\rangle\otimes|0\rangle$
and the "entangled"  $(|0\rangle\otimes|1\rangle-|1\rangle\otimes|0\rangle)\otimes|1\rangle$
belongs to $\Lambda({\cal H}_3(2)).$
This kind of puzzle  is solved by observing that
the  entanglement of, say $|a\rangle\otimes|b\rangle- |b\rangle\otimes|a\rangle,$ 
is not {\em physical}. Indeed  in  the involved subsystems i.e., individual "labelled" particles,
due to the very notion of indistinguishability,
are physically not {\em accessible}.

This situation is just an illustration  of the
relativity of the notion of entanglement \cite{virtual}.
The latter crucially 
 depends on the choice of a particular partition into physical subsystems.
In this case  "good" subsystems are associated with the set of single particle modes 
(labelled by $l\in{\bf{N}}_L$ whose occupation numbers are   physical observables and {\em not}
with the particles themselves.
From this perspective one can have entanglement {\em without} entanglement.
For instance  a {\em one}-particle state e.g., $|0\rangle\otimes|1\rangle+|1\rangle\otimes|0\rangle,$
can be -- with respect to the  partition into mode subsystems--   entangled.
It is important to stress that such kind of one-particle  entanglement (or correlation),
despite its paradoxical nature, has been recently proven to allow for  quantum teleportation
\cite{Lee};  therefore it has to be regarded as a genuine {\em resource} for QIP. 

The  Fock space ${\cal H}_L,$ since it allows for a varying particle occupation,
does  {\em not} correspond generally to the state space of a physical system. 
For charged fermions coherent superpositions of vectors belonging to different
particle number sectors are forbidden due to  charge superselection rule \cite{selection}.
In this sense
our qubits are {\em unphysical}. Only qubit states in the $\Lambda({\cal H}_L(N))$
are associated with ($N$-particle) physical states.
Accordingly not all the elements  of End$({\cal H}_L)$ correspond to  physical observables:
the latter span  the subalgebra  $\cal F$ of   number conserving operators i.e.,
%\begin{equation}
${\cal F}:= \{X\,/\,[X,\,N]=0\}=\oplus_N \mbox{End}({\cal H}_L(N))
$
%\label{observables}
%\end{equation}

{\em Local Entanglement.}%%%%%%%%%%%%%%%%%%%%%%%%%%%%%%%%%%%%%%%%%%%%%%%%%%%%%%%%%%%%
Let $|\Psi\rangle\in{\cal H}_L(N)$ be the associated $j$-th {\em local} density matrix is given by
$\rho_j:= \mbox{Tr}_{\underline{j}} |\Psi\rangle\langle\Psi|$ where 
$ \mbox{Tr}_{\underline{j}}$ denotes the trace over all but the $j$-th sites.
For any $j\in{\bf{N}}_L$ one obtains  a
 bipartition of ${\cal H}_L$ i.e., $\CC^2\otimes (\CC^2)^{\otimes\,(L-1)}$
therefore the entropy $S$ (von Neumann as well as linear) of $\rho_j$ is a measure of the entanglement 
of the $j$-th site with the remaining $N-1$ ones.

Local entanglement is {\em relative} to the decomposition into subsystems
i.e., sites, defined by  mapping (\ref{mapping}) \cite{virtual}.
One could consider different isomorphisms
giving rise to {\em inequivalent}  partitions  in to "local" subsystems.
This fact can be clearly seen by introducing 
creation and annihilation operators
$\{c_j\}_{j=1}^L\subset \mbox{End}({\cal H}_L),  $
( $[a,\,b]_\pm= a b\pm b a$) 
which satisfy  canonical (anti) commutation relations for (fermions) bosons
\begin{equation}
[ c_i,\,c_j]_\pm=0,\;
[c_i,\,c_j^\dagger]_\pm= \delta_{ij},\; c_j\,|0\rangle=0 \, (j\in{\bf{N}}_L).
\label{car}
\end{equation}
Of course 
${\cal H}_L=\mbox{span} \{ \prod_{j=1}^L (c_j^\dagger)^{n_j}|0\rangle
\,/\, n_1,\ldots,n_L\in{\bf{N}}_\infty\}.$
If $U$ is a $L\times L$ unitary matrix then it is well-known that 
the following (Bogoliubov) transformation
\begin{equation}
c_i\rightarrow  \tilde c_i:=\sum_{j=1}^L U_{ij}\,c_j, \qquad(i\in{\bf{N}}_L) 
\label{bogo}
\end{equation}
maps  fermions  ( bosons) onto fermions (boson) 
giving rise to an automorphism of the  observable algebra.
Accordingly   new  occupation-number representations 
$\Lambda_U\colon \prod_{i=1}^L  ({\tilde c}_i^\dagger)^{n_i}\, |0\rangle\mapsto
\otimes_{i=1}^L |n_i\rangle\,(n_i=0,1)$
are  defined. Clearly entanglement is strongly {\em relative}
to the decompositions  associated with different $\Lambda_U$'s.
Notice that even though  automorphsims     
(\ref{bogo}) have a  single-particle origin,  
they define {\em non-local} transformations of the Fock space onto itself.
Indeed a mapping $W\in{\cal U}(({\cal H}_L)$ is {\em local}
with respect the subsystem decomposition associated with $\Lambda_U$
iff $\Lambda_U\circ W\circ \Lambda^{-1}_U\in \prod_{i=1}^L {\cal U}(\CC^2)_i$

As a particular, though  quite relevant, case one can consider the Fourier transformation
i.e., $U_{kj}:=e^{i\,kj},\,k:= 2\pi(l-1)/L, (l\in{\bf{N}}_L).$
The wave-vectors $k$ label the so-called reciprocal  lattice 
($\Lambda_U$ is denoted by $\Lambda^*$) and represent physically modes delocalized
over the spatial lattice. 
It is obvious that states that are entangled (non-entangled)  with respect $\Lambda$ 
can be non non-entangled (entangled) with respect $\Lambda^*.$

The situation we shall investigate in this paper is the following.
Suppose $H\in\mbox{End}({\cal H}_L)$ is a { non degenerate} (gran-canonical) hamiltonian
and $H=\sum_{m}  \epsilon_m \,| \epsilon_m\rangle\langle  \epsilon_m|$ its spectral decomposition.
If $\rho_j^m$ denotes the $j$-th local density matrix associated with the energy eigenstate
$\varepsilon_m,$ one can  compute the  quantity 
\begin{equation}
S_{1/\beta} := \frac{1}{Z\,L}\sum_{m=1}^{2^L} e^{-\beta \epsilon_m} \sum_{j=1}^L S(\rho_j^m) 
\label{aim}
\end{equation}
where $Z$
is the (gran-canonical) partition function i.e., $Z:= \sum_{m=1}^{2^L} e^{-\beta \epsilon_m}.$
Eq. (\ref{aim}) is the thermal expectation value of the local entanglement averaged over the whole 
lattice \cite{not}.
In particular we will be interested in  the limit $\beta\mapsto\infty$
i.e.,  local entanglement $S_0$ in the {\em ground state}.   

When energy spectrum shows degeneracies
Eq (\ref{aim}) is no longer well-defined. We assume that
there  is a "natural" (see examples below)
way to select a a complete set of commuting   observables
containing $H,$ whose joint  eigenvectors
provide the $\epsilon_m$'s to be used in (\ref{aim}).

To begin with
we  observe  that
\begin{equation}
\rho_i= |1\rangle\langle 1| \,\langle \Psi|n_i|\Psi\rangle +  |0\rangle\langle 0|
 \,\langle\Psi|\openone - n_i|\Psi\rangle
\label{rhoi}
\end{equation}
where $n_i :=c_i^\dagger\,c_i =|1\rangle\langle1|_i \otimes\openone_{\underline{i}}$
is the local occupation number projector.
Indeed:
$\langle 1|\rho_i|1\rangle=\mbox{Tr} ( |0\rangle\langle 0|\,\rho_i)=\mbox{Tr} (n_i\,\rho_i)=
\langle \Psi|n_i|\Psi\rangle$
in the same way one obtains the other diagonal element of $\rho_i.$
Moreover
$\langle 0|\rho_i|1\rangle=\mbox{Tr} ( |1\rangle\langle 0|\,\rho_i)= \mbox{Tr} (c^\dagger_i\,\rho_i)=
\langle \Psi|c^\dagger_i|\Psi\rangle=0,$
 last equality is due to the fact that $|\Psi\rangle$ is a particle number eigenstate
i.e., an eigenstate of the operator $\hat N:=\sum_{j=}^L n_j.$

{\em Itinerant  Fermions.}
We now consider  free (spinless)  fermions hopping 
in the lattice. 
The Hamiltonian is given by
\begin{equation}
H_{Free}=-t\, \sum_{j=1}^{L-1} (c_{j+1}^\dagger \,c_j+\mbox{h.c.})-\mu\hat N 
\label{free}
\end{equation}
Introducing the Fourier  fermionic operators
$c_k:= 1/\sqrt{L}\sum_{j=1}^L e^{i\,k\,j} c_j,$ 
it is a text-book exercise to  prove that  (\ref{free}) 
has eigenstates given by the $N$-particle vectors
$|{\bf{k}}\rangle:= \prod_{m=1}^L c_{k_m}^\dagger |0\rangle,\,({\bf{k}}:=(k_1,\ldots,k_N)\in{\bf{R}}^N$
with eigenvalues $\epsilon_{\bf{k}} := -2\,t \sum_{m=1}^N \cos(k_m)-\mu N.$ 

The local  density matrix is easily obtained by using Eq. (\ref{rhoi})
and the translational  properties of the $|{\bf{k}}\rangle$'s.
If $T$ denotes the natural representation in ${\cal H}_L$ of the cyclic permutations $i\mapsto i+1$ mod $L$
i.e., the translation operator, one has 
$T\,|{\bf{k}}\rangle= e^{i\sum_{m=1}^N k_m}\,|{\bf{k}}\rangle.$
Therefore
%\begin{equation}
$\langle{\bf{k}}| n_j |{\bf{k}}\rangle ={1}/{L} \sum_{i=1}^L \langle{\bf{k}}| n_i |{\bf{k}}\rangle=
N/L=:n .$
%\end{equation}
Whereby 
$
E= 1/{Z} \sum_{N=0}^L S(N/L) e^{\beta\mu N}\, Z_N(\beta)= {1}/{Z}\,
\mbox{Tr}\, ( S(\hat N/L)\, {e^{-\beta\,H_{Free}}}) ,
$
in which $S(n)= -n\,\ln n -(1-n)\,\ln (1-n)$  and
$Z_N(\beta):=\mbox{Tr}_{{\cal H}_L(N)} e^{-\beta \,(H + \mu \hat N)}$
is the ($N$-particle)  canonical partition function.

The fraction $p(N):=e^{\beta\mu N}\, Z_N/Z$ gives of course the probability
of having  any $N$-particle configuration. In the thermodynamical  limit ($N,\,L\mapsto\infty, N/L=$const)
 $p(N)$ becomes strongly peaked around the expectation value $N_0$ of $\hat N.$
In this case local entanglement is simply given by the Shannon function $E\sim S(n_0),$ it readily displays 
an intuitive feature:
local entanglement vanishes  for the empty (fully filled) lattice
being the unique associated state given the product   $|0\rangle$ ($\otimes_{l}|1\rangle_l$);
moreover $E$ is maximal at half-filling i.e., $n_0=1/2$.
Notice  that for the states $|{\bf{k}}\rangle$  entanglement associated with the 
$\Lambda^*$ partition is obviously zero.

{\em Spin $1/2$ Fermions.}
Here we consider the  lattice models of spin $1/2$ fermion  model. We have then to introduce an extra
dicothomic variable  $\sigma=\uparrow,\downarrow$ to label   the single- particle state-vectors.
As usual fermionic operators corresponding to different $\sigma$'s always anti-commute.
In this case it is convenient to consider the  $2^{2\,L}$-dimensional
Fock space as  isomorphic teh $L$-fold tensor power of  four-dimensional
space i.e, ${\cal H}_F\cong (\CC^4)^{\otimes\,L}.$
The  
local  state space is spanned by  the vacuum $|0\rangle$ and the vectors 
\begin{equation}
|\uparrow\rangle_j
:=c^\dagger_{j\uparrow} |0\rangle,\,
 |\downarrow\rangle_j:=c^\dagger_{j\downarrow} |0\rangle,\, |\uparrow\downarrow\rangle_j
:= c^\dagger_{j\downarrow} 
c^\dagger_{j\uparrow} |0\rangle.
\end{equation}

The $\rho_j=\mbox{Tr}_{\underline{j}} |\Psi\rangle\langle\Psi|$
is now a $4\times 4 $ matrix. If the  $N$-particle state $|\Psi\rangle$
is a) translational invariant, b) eigenstate  of $S^z:=\sum_{j=1}^L (n_{j\uparrow} -n_{j\downarrow}),$
is easy to see that
%\begin{equation}
$\rho_j=1/L\, \mbox{diag} ( 1- N_\uparrow - N_\downarrow-N_l,\, N_\uparrow,\, N_\downarrow,\, N_l)
$
%\end{equation}
where $N_\sigma:= \sum_{j=1}^L \langle\Psi|  n_{j\sigma}(1-n_{j-\sigma})|\Psi\rangle, 
\,(\sigma=\uparrow, \downarrow)$ is the number of lattice sites singly occupied by a
$\sigma$ fermion
and $N_l:= \sum_{j=1}^L \langle\Psi|  n_{j\uparrow} n_{j\downarrow} |\Psi\rangle$
is the number of doubly occupied sites.
We see that local entanglement, in  state $|\Psi\rangle$
is a function just of  the occupation numbers $N_\alpha,\,(\alpha=\uparrow, \downarrow, l),$
in particular it follows that Eq. (\ref{aim}) can be effectively  computed 
for Hamiltonians commuting with the $N_\alpha$'s
i.e., $E=\sum_{ \{N_\alpha\} } e^{\beta\mu N} S( \{N_\alpha\})\,Z( \{N_\alpha\})/Z.$
An  instance of this case is illustrated in the following.

{\em Supersymmetric Dimer.}
We consider here a  two-site i.e, a {\em dimer}, version of the so called supersymmetric EKS model
\cite{EKS}. For zero chemical potential i.e., half-filling the  EKS 
Hamiltonian acts on the basis states as follows
\begin{equation}
H\,|\alpha\rangle\otimes|\beta\rangle= (-1)^{|\alpha|\,|\beta|}\,|\beta\rangle\otimes|\alpha\rangle 
\label{eks}
\end{equation}
where $|\alpha|$ is the {\em parity} of the single-site state $|\alpha\rangle$
i.e., $|\uparrow|=|\downarrow|=1,\, |0|=|\uparrow\downarrow|=0.$ 
Since $H$ is just a {\em graded} permutator the relations
$[H,\,N_\alpha]=0$  hold true.
The state-space split according the $N_\alpha$'s configurations
${\cal H}_L= \oplus_{\{N_\alpha\}} {\cal H}(\{N_\alpha\})$
and the Hamiltonian can be  diagonalized within each sector.
Notice that Eq. (\ref{eks}) is also invariant under a global 
{\em particle-hole} transformation i.e., $|\sigma\rangle\leftrightarrow |-\sigma\rangle,$
$|0\rangle\leftrightarrow |\uparrow\downarrow\rangle.$

It is straightforward to check that $H$ admits four singlets non-entangled
(the configurations $(0,0,0),\,(2,0,0)$ along with their particle-hole conjugates)
and six doublets ( $(1,0,0),\,(0,1,0)$ and conjugates and the self-conjugated
$(1,1,0),\,(0,0,1)$) with entanglement $\ln 2.$ Moreover since $H^2=\openone$ one gets
an energy spectrum given by $\{-1,\,1\},$ being both the eigenvalues $8$-fold degenerated.     
Therefore 
$E= 12\,\ln 2\,\cosh\beta/16\,\cosh\beta=3/4\,\ln 2:$
the local-entanglement (at half filling) is temperature independent. 

This very simple result is due to the large symmetry group
of the Hamiltonian (\ref{eks}). A more interesting case is obtained 
introducing a model in which  a free parameter controls
the competition between the localized and itinerant  nature of the 
lattice fermions. 

{\em Hubbard Dimer.}
If $H_{Free}^\sigma$  simply denotes Eq. (\ref{free}) with extra spin index then 
 Hubbard Hamiltonian  reads (\ref{Hubb})
\begin{eqnarray}
H_{Hubb}= \sum_{\sigma=\uparrow,\downarrow} H_{Free}^\sigma 
+U \sum_{j=1}^L  n_{j\uparrow} n_{j\downarrow}. 
\label{Hubb}
\end{eqnarray}
The new local terms added account for the on-site interaction e.g., Coulomb repulsion,
experienced  by pairs of (opposite) spin fermions sitting on the same lattice site. 
By introducing the total  fixed spin  number operators
$\hat N_{\sigma}:= \sum_{j=1}^L n_{j\sigma}\,(\sigma=\uparrow,\downarrow)$
is easy to check that both of them commute with the Hubbard Hamiltonian (\ref{Hubb}).
This implies that $H_{Hubb}$ can be separately diagonalized
in each joint eigenspace  ${\cal H}(N_\uparrow, N_\downarrow)$ of the $\hat N_\sigma$'s.
In the  the dimer  case i.e, the {\em dimer}, one finds 
dim ${\cal H}(N_\uparrow, N_\downarrow)=\prod_\sigma \pmatrix{2\cr N_\sigma};$
then at most (for $N_\uparrow=N_\downarrow=1$) one has to solve 
a four-dimensional diagonalization problem.
%%%%%%%%%%%%%%%
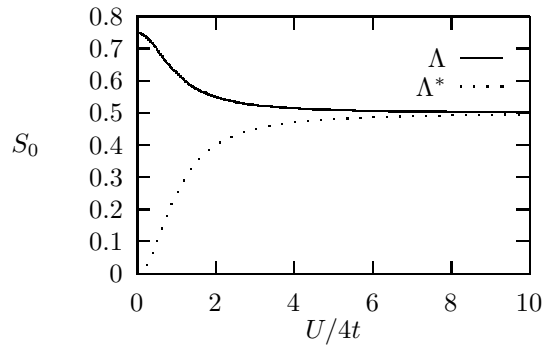
\begin{figure}
\begin{center}
\input{s0.tex}
\end{center}
\caption{
Entanglement of the Hubbard dimer ground state as a function of $U/4\,t
$ for decomposition associated with real and reciprocal lattice.
}
\protect\label{Fig1}
\end{figure}
%%%%%%%%%%%%%%%%%%%%%%%%%%%

The (unnormalized) ground state, for the repulsive case $U>0,$  is given by
$-G_0\,|0\rangle$ where
\begin{equation}
G_0:= c_{1\uparrow}^\dagger\,c_{1\downarrow}^\dagger+ c_{2\uparrow}^\dagger\,c_{2\downarrow}^\dagger
+\alpha_+(U/4t)\,( c_{1\uparrow}^\dagger\,c_{2\downarrow}^\dagger -  c_{1\downarrow}^\dagger\,c_{2\uparrow}^\dagger)
\label{hub_gs}
\end{equation}
where $\alpha_\pm(x) := x\pm \sqrt{1+x^2},$ and the associated eigenvalue is given by
$E_0=-2 t\,\alpha_-.$
The entanglement of the state (\ref{hub_gs}) is easily studied as a function $U/4t.$
Using  liner entropy as an entanglement measure one finds
$S_0(U/t)=1-\mbox{Tr}\,\rho_0^2=1-1/2 (\alpha_+^4+1)\,(\alpha_+^2+1)^{-2}.$
Local entanglement  is monotonically decreasing as a function of $U/4\,t.$
(Fig. 1).
In particular one obtains
the free limit $S_0(0)=3/4$ and the strong coupling limit $S_0(\infty)=1/2$
that correspond to ground states given by uniform superpositions of
respectively four and two states [see Eq.(\ref{hub_gs}].
Of course the physical interpretation is quite simple:
the higher  the on-site repulsion $U$ the more
local charge fluctuations are suppressed and 
the smaller the number of available states.
Eventually for infinite repulsion  doubly occupied sites
get  decoupled and only spin fluctuations survive.
In this regime  Hubbard model is known to be equivalent to an
{\em anti}-ferromagnetic Heisenberg model for spin $1/2$ \cite{eq_heis}.
Ground state as well as thermal entanglement for this (and related)
models have been quite recently studied \cite{thermal}.

It is instructive to write the dimer ground state creator (\ref{hub_gs})
in terms of the Fourier  operators $c^{(\pm)}_\sigma := 1/\sqrt{2}\,(c_{1\sigma}\pm  c_{2\sigma})\, 
(\sigma=\uparrow,\downarrow),$ from Eq (\ref{hub_gs}) one finds
\begin{eqnarray}
G_0(\alpha) = \sum_{k=\pm}( 1+k\,\alpha_+)\, c_\uparrow^{(k)\dagger }\, c_\downarrow^{(k)\dagger}.
\end{eqnarray}
With respect this reciprocal  decomposition local entanglement 
is   an {\em increasing}  function of $U/4 t.$
From the free case ($\alpha_+=1$),
that is unentangled up to strong coupling ($\alpha_+=\infty$) 
which gives $S_0^*(\infty)=1/2$ (see Fig. 1).

The example of the Hubbard dimer shows that -not surprisingly -  entanglement is well-suited 
to analyze the interplay
between itinerant and  localized  features of Hubbard Model (\ref{Hubb}):
hopping term $t$ (repulsion $U$ ) term  is responsible for entanglement in the real
(reciprocal) lattice decomposition.

{\em Conclusions.}
In this paper we discussed some issues related to  entanglement
in system of indistinguishable particle.
For these systems  quantum statistics applies 
and therefore their state-space is not naturally endowed
with a tensor product structure.

Nevertheless  mappings between their Fock spaces and multi-partite
state-spaces can be established (the well-known occupation number representation)
and then  the usual definition of entanglement can be applied.
For systems with $L$ single particle states available
the set of possible inequivalent decompositions into $L$ subsystems (modes)
is  parametrized by the group $U(L)$ of Bogoliubov transformations.

We focused on simple e.g., Hubbard, models of fermions on lattice
studying how, as function of the model parameters,  local 
entanglement varies both with respect to  real and reciprocal lattice decomposition. 
Results suggest  that this   notion of entanglement is well suited to describe
interplay between localization and itinerancy  in  these systems. 

We believe that  the approach pursued
in this paper -- besides to establish   a connection between the field
of Quantum Information Processing  and condensed matter physics --
can  provide novel physical insights
in the study of interacting ensembles of indistinguishable particles.

I acknowledge  discussions with R. R. Zapatrin, Lara Faoro and X-G. Wang.
Moreover I thank the authors of Ref. \cite{john} for drawing my attention on their work
and for stimulating correspondence

\end{multicols}%%%%%%%%%%%%%%%%%%%%%%%%%%%%%%%%%%%%%%%%%%%%
\end{document}

%% file: s0.tex
% GNUPLOT: LaTeX picture
\setlength{\unitlength}{0.240900pt}
\ifx\plotpoint\undefined\newsavebox{\plotpoint}\fi
\begin{picture}(900,540)(0,0)
\font\gnuplot=cmr10 at 10pt
\gnuplot
\sbox{\plotpoint}{\rule[-0.200pt]{0.400pt}{0.400pt}}%
\put(220.0,113.0){\rule[-0.200pt]{148.394pt}{0.400pt}}
\put(220.0,113.0){\rule[-0.200pt]{0.400pt}{97.324pt}}
\put(220.0,113.0){\rule[-0.200pt]{4.818pt}{0.400pt}}
\put(198,113){\makebox(0,0)[r]{0}}
\put(816.0,113.0){\rule[-0.200pt]{4.818pt}{0.400pt}}
\put(220.0,164.0){\rule[-0.200pt]{4.818pt}{0.400pt}}
\put(198,164){\makebox(0,0)[r]{0.1}}
\put(816.0,164.0){\rule[-0.200pt]{4.818pt}{0.400pt}}
\put(220.0,214.0){\rule[-0.200pt]{4.818pt}{0.400pt}}
\put(198,214){\makebox(0,0)[r]{0.2}}
\put(816.0,214.0){\rule[-0.200pt]{4.818pt}{0.400pt}}
\put(220.0,265.0){\rule[-0.200pt]{4.818pt}{0.400pt}}
\put(198,265){\makebox(0,0)[r]{0.3}}
\put(816.0,265.0){\rule[-0.200pt]{4.818pt}{0.400pt}}
\put(220.0,315.0){\rule[-0.200pt]{4.818pt}{0.400pt}}
\put(198,315){\makebox(0,0)[r]{0.4}}
\put(816.0,315.0){\rule[-0.200pt]{4.818pt}{0.400pt}}
\put(220.0,366.0){\rule[-0.200pt]{4.818pt}{0.400pt}}
\put(198,366){\makebox(0,0)[r]{0.5}}
\put(816.0,366.0){\rule[-0.200pt]{4.818pt}{0.400pt}}
\put(220.0,416.0){\rule[-0.200pt]{4.818pt}{0.400pt}}
\put(198,416){\makebox(0,0)[r]{0.6}}
\put(816.0,416.0){\rule[-0.200pt]{4.818pt}{0.400pt}}
\put(220.0,467.0){\rule[-0.200pt]{4.818pt}{0.400pt}}
\put(198,467){\makebox(0,0)[r]{0.7}}
\put(816.0,467.0){\rule[-0.200pt]{4.818pt}{0.400pt}}
\put(220.0,517.0){\rule[-0.200pt]{4.818pt}{0.400pt}}
\put(198,517){\makebox(0,0)[r]{0.8}}
\put(816.0,517.0){\rule[-0.200pt]{4.818pt}{0.400pt}}
\put(220.0,113.0){\rule[-0.200pt]{0.400pt}{4.818pt}}
\put(220,68){\makebox(0,0){0}}
\put(220.0,497.0){\rule[-0.200pt]{0.400pt}{4.818pt}}
\put(343.0,113.0){\rule[-0.200pt]{0.400pt}{4.818pt}}
\put(343,68){\makebox(0,0){2}}
\put(343.0,497.0){\rule[-0.200pt]{0.400pt}{4.818pt}}
\put(466.0,113.0){\rule[-0.200pt]{0.400pt}{4.818pt}}
\put(466,68){\makebox(0,0){4}}
\put(466.0,497.0){\rule[-0.200pt]{0.400pt}{4.818pt}}
\put(590.0,113.0){\rule[-0.200pt]{0.400pt}{4.818pt}}
\put(590,68){\makebox(0,0){6}}
\put(590.0,497.0){\rule[-0.200pt]{0.400pt}{4.818pt}}
\put(713.0,113.0){\rule[-0.200pt]{0.400pt}{4.818pt}}
\put(713,68){\makebox(0,0){8}}
\put(713.0,497.0){\rule[-0.200pt]{0.400pt}{4.818pt}}
\put(836.0,113.0){\rule[-0.200pt]{0.400pt}{4.818pt}}
\put(836,68){\makebox(0,0){10}}
\put(836.0,497.0){\rule[-0.200pt]{0.400pt}{4.818pt}}
\put(220.0,113.0){\rule[-0.200pt]{148.394pt}{0.400pt}}
\put(836.0,113.0){\rule[-0.200pt]{0.400pt}{97.324pt}}
\put(220.0,517.0){\rule[-0.200pt]{148.394pt}{0.400pt}}
\put(45,315){\makebox(0,0){$S_0$}}
\put(528,23){\makebox(0,0){$U/4t$}}
\put(220.0,113.0){\rule[-0.200pt]{0.400pt}{97.324pt}}
\put(706,452){\makebox(0,0)[r]{$\Lambda$}}
\put(728.0,452.0){\rule[-0.200pt]{15.899pt}{0.400pt}}
\put(220,492){\usebox{\plotpoint}}
\put(220,490.67){\rule{0.723pt}{0.400pt}}
\multiput(220.00,491.17)(1.500,-1.000){2}{\rule{0.361pt}{0.400pt}}
\put(226,489.17){\rule{0.700pt}{0.400pt}}
\multiput(226.00,490.17)(1.547,-2.000){2}{\rule{0.350pt}{0.400pt}}
\put(229,487.17){\rule{0.700pt}{0.400pt}}
\multiput(229.00,488.17)(1.547,-2.000){2}{\rule{0.350pt}{0.400pt}}
\multiput(232.00,485.95)(0.462,-0.447){3}{\rule{0.500pt}{0.108pt}}
\multiput(232.00,486.17)(1.962,-3.000){2}{\rule{0.250pt}{0.400pt}}
\multiput(235.00,482.95)(0.462,-0.447){3}{\rule{0.500pt}{0.108pt}}
\multiput(235.00,483.17)(1.962,-3.000){2}{\rule{0.250pt}{0.400pt}}
\multiput(238.00,479.95)(0.685,-0.447){3}{\rule{0.633pt}{0.108pt}}
\multiput(238.00,480.17)(2.685,-3.000){2}{\rule{0.317pt}{0.400pt}}
\multiput(242.61,475.37)(0.447,-0.685){3}{\rule{0.108pt}{0.633pt}}
\multiput(241.17,476.69)(3.000,-2.685){2}{\rule{0.400pt}{0.317pt}}
\multiput(245.61,471.37)(0.447,-0.685){3}{\rule{0.108pt}{0.633pt}}
\multiput(244.17,472.69)(3.000,-2.685){2}{\rule{0.400pt}{0.317pt}}
\multiput(248.00,468.95)(0.462,-0.447){3}{\rule{0.500pt}{0.108pt}}
\multiput(248.00,469.17)(1.962,-3.000){2}{\rule{0.250pt}{0.400pt}}
\multiput(251.61,463.82)(0.447,-0.909){3}{\rule{0.108pt}{0.767pt}}
\multiput(250.17,465.41)(3.000,-3.409){2}{\rule{0.400pt}{0.383pt}}
\multiput(254.61,459.37)(0.447,-0.685){3}{\rule{0.108pt}{0.633pt}}
\multiput(253.17,460.69)(3.000,-2.685){2}{\rule{0.400pt}{0.317pt}}
\multiput(257.61,455.37)(0.447,-0.685){3}{\rule{0.108pt}{0.633pt}}
\multiput(256.17,456.69)(3.000,-2.685){2}{\rule{0.400pt}{0.317pt}}
\multiput(260.61,451.37)(0.447,-0.685){3}{\rule{0.108pt}{0.633pt}}
\multiput(259.17,452.69)(3.000,-2.685){2}{\rule{0.400pt}{0.317pt}}
\multiput(263.61,447.37)(0.447,-0.685){3}{\rule{0.108pt}{0.633pt}}
\multiput(262.17,448.69)(3.000,-2.685){2}{\rule{0.400pt}{0.317pt}}
\multiput(266.61,443.37)(0.447,-0.685){3}{\rule{0.108pt}{0.633pt}}
\multiput(265.17,444.69)(3.000,-2.685){2}{\rule{0.400pt}{0.317pt}}
\multiput(269.00,440.95)(0.462,-0.447){3}{\rule{0.500pt}{0.108pt}}
\multiput(269.00,441.17)(1.962,-3.000){2}{\rule{0.250pt}{0.400pt}}
\multiput(272.61,436.37)(0.447,-0.685){3}{\rule{0.108pt}{0.633pt}}
\multiput(271.17,437.69)(3.000,-2.685){2}{\rule{0.400pt}{0.317pt}}
\multiput(275.00,433.95)(0.685,-0.447){3}{\rule{0.633pt}{0.108pt}}
\multiput(275.00,434.17)(2.685,-3.000){2}{\rule{0.317pt}{0.400pt}}
\multiput(279.00,430.95)(0.462,-0.447){3}{\rule{0.500pt}{0.108pt}}
\multiput(279.00,431.17)(1.962,-3.000){2}{\rule{0.250pt}{0.400pt}}
\multiput(282.00,427.95)(0.462,-0.447){3}{\rule{0.500pt}{0.108pt}}
\multiput(282.00,428.17)(1.962,-3.000){2}{\rule{0.250pt}{0.400pt}}
\multiput(285.00,424.95)(0.462,-0.447){3}{\rule{0.500pt}{0.108pt}}
\multiput(285.00,425.17)(1.962,-3.000){2}{\rule{0.250pt}{0.400pt}}
\multiput(288.00,421.95)(0.462,-0.447){3}{\rule{0.500pt}{0.108pt}}
\multiput(288.00,422.17)(1.962,-3.000){2}{\rule{0.250pt}{0.400pt}}
\multiput(291.00,418.95)(0.462,-0.447){3}{\rule{0.500pt}{0.108pt}}
\multiput(291.00,419.17)(1.962,-3.000){2}{\rule{0.250pt}{0.400pt}}
\put(294,415.17){\rule{0.700pt}{0.400pt}}
\multiput(294.00,416.17)(1.547,-2.000){2}{\rule{0.350pt}{0.400pt}}
\multiput(297.00,413.95)(0.462,-0.447){3}{\rule{0.500pt}{0.108pt}}
\multiput(297.00,414.17)(1.962,-3.000){2}{\rule{0.250pt}{0.400pt}}
\put(300,410.17){\rule{0.700pt}{0.400pt}}
\multiput(300.00,411.17)(1.547,-2.000){2}{\rule{0.350pt}{0.400pt}}
\put(303,408.17){\rule{0.700pt}{0.400pt}}
\multiput(303.00,409.17)(1.547,-2.000){2}{\rule{0.350pt}{0.400pt}}
\put(306,406.17){\rule{0.700pt}{0.400pt}}
\multiput(306.00,407.17)(1.547,-2.000){2}{\rule{0.350pt}{0.400pt}}
\put(309,404.17){\rule{0.700pt}{0.400pt}}
\multiput(309.00,405.17)(1.547,-2.000){2}{\rule{0.350pt}{0.400pt}}
\put(312,402.67){\rule{0.723pt}{0.400pt}}
\multiput(312.00,403.17)(1.500,-1.000){2}{\rule{0.361pt}{0.400pt}}
\put(315,401.17){\rule{0.900pt}{0.400pt}}
\multiput(315.00,402.17)(2.132,-2.000){2}{\rule{0.450pt}{0.400pt}}
\put(319,399.17){\rule{0.700pt}{0.400pt}}
\multiput(319.00,400.17)(1.547,-2.000){2}{\rule{0.350pt}{0.400pt}}
\put(322,397.67){\rule{0.723pt}{0.400pt}}
\multiput(322.00,398.17)(1.500,-1.000){2}{\rule{0.361pt}{0.400pt}}
\put(325,396.67){\rule{0.723pt}{0.400pt}}
\multiput(325.00,397.17)(1.500,-1.000){2}{\rule{0.361pt}{0.400pt}}
\put(328,395.17){\rule{0.700pt}{0.400pt}}
\multiput(328.00,396.17)(1.547,-2.000){2}{\rule{0.350pt}{0.400pt}}
\put(331,393.67){\rule{0.723pt}{0.400pt}}
\multiput(331.00,394.17)(1.500,-1.000){2}{\rule{0.361pt}{0.400pt}}
\put(334,392.67){\rule{0.723pt}{0.400pt}}
\multiput(334.00,393.17)(1.500,-1.000){2}{\rule{0.361pt}{0.400pt}}
\put(337,391.67){\rule{0.723pt}{0.400pt}}
\multiput(337.00,392.17)(1.500,-1.000){2}{\rule{0.361pt}{0.400pt}}
\put(340,390.67){\rule{0.723pt}{0.400pt}}
\multiput(340.00,391.17)(1.500,-1.000){2}{\rule{0.361pt}{0.400pt}}
\put(343,389.67){\rule{0.723pt}{0.400pt}}
\multiput(343.00,390.17)(1.500,-1.000){2}{\rule{0.361pt}{0.400pt}}
\put(346,388.67){\rule{0.723pt}{0.400pt}}
\multiput(346.00,389.17)(1.500,-1.000){2}{\rule{0.361pt}{0.400pt}}
\put(349,387.67){\rule{0.723pt}{0.400pt}}
\multiput(349.00,388.17)(1.500,-1.000){2}{\rule{0.361pt}{0.400pt}}
\put(352,386.67){\rule{0.964pt}{0.400pt}}
\multiput(352.00,387.17)(2.000,-1.000){2}{\rule{0.482pt}{0.400pt}}
\put(356,385.67){\rule{0.723pt}{0.400pt}}
\multiput(356.00,386.17)(1.500,-1.000){2}{\rule{0.361pt}{0.400pt}}
\put(223.0,491.0){\rule[-0.200pt]{0.723pt}{0.400pt}}
\put(362,384.67){\rule{0.723pt}{0.400pt}}
\multiput(362.00,385.17)(1.500,-1.000){2}{\rule{0.361pt}{0.400pt}}
\put(365,383.67){\rule{0.723pt}{0.400pt}}
\multiput(365.00,384.17)(1.500,-1.000){2}{\rule{0.361pt}{0.400pt}}
\put(359.0,386.0){\rule[-0.200pt]{0.723pt}{0.400pt}}
\put(371,382.67){\rule{0.723pt}{0.400pt}}
\multiput(371.00,383.17)(1.500,-1.000){2}{\rule{0.361pt}{0.400pt}}
\put(374,381.67){\rule{0.723pt}{0.400pt}}
\multiput(374.00,382.17)(1.500,-1.000){2}{\rule{0.361pt}{0.400pt}}
\put(368.0,384.0){\rule[-0.200pt]{0.723pt}{0.400pt}}
\put(380,380.67){\rule{0.723pt}{0.400pt}}
\multiput(380.00,381.17)(1.500,-1.000){2}{\rule{0.361pt}{0.400pt}}
\put(377.0,382.0){\rule[-0.200pt]{0.723pt}{0.400pt}}
\put(386,379.67){\rule{0.723pt}{0.400pt}}
\multiput(386.00,380.17)(1.500,-1.000){2}{\rule{0.361pt}{0.400pt}}
\put(383.0,381.0){\rule[-0.200pt]{0.723pt}{0.400pt}}
\put(392,378.67){\rule{0.964pt}{0.400pt}}
\multiput(392.00,379.17)(2.000,-1.000){2}{\rule{0.482pt}{0.400pt}}
\put(389.0,380.0){\rule[-0.200pt]{0.723pt}{0.400pt}}
\put(402,377.67){\rule{0.723pt}{0.400pt}}
\multiput(402.00,378.17)(1.500,-1.000){2}{\rule{0.361pt}{0.400pt}}
\put(396.0,379.0){\rule[-0.200pt]{1.445pt}{0.400pt}}
\put(408,376.67){\rule{0.723pt}{0.400pt}}
\multiput(408.00,377.17)(1.500,-1.000){2}{\rule{0.361pt}{0.400pt}}
\put(405.0,378.0){\rule[-0.200pt]{0.723pt}{0.400pt}}
\put(417,375.67){\rule{0.723pt}{0.400pt}}
\multiput(417.00,376.17)(1.500,-1.000){2}{\rule{0.361pt}{0.400pt}}
\put(411.0,377.0){\rule[-0.200pt]{1.445pt}{0.400pt}}
\put(429,374.67){\rule{0.964pt}{0.400pt}}
\multiput(429.00,375.17)(2.000,-1.000){2}{\rule{0.482pt}{0.400pt}}
\put(420.0,376.0){\rule[-0.200pt]{2.168pt}{0.400pt}}
\put(442,373.67){\rule{0.723pt}{0.400pt}}
\multiput(442.00,374.17)(1.500,-1.000){2}{\rule{0.361pt}{0.400pt}}
\put(433.0,375.0){\rule[-0.200pt]{2.168pt}{0.400pt}}
\put(454,372.67){\rule{0.723pt}{0.400pt}}
\multiput(454.00,373.17)(1.500,-1.000){2}{\rule{0.361pt}{0.400pt}}
\put(445.0,374.0){\rule[-0.200pt]{2.168pt}{0.400pt}}
\put(473,371.67){\rule{0.723pt}{0.400pt}}
\multiput(473.00,372.17)(1.500,-1.000){2}{\rule{0.361pt}{0.400pt}}
\put(457.0,373.0){\rule[-0.200pt]{3.854pt}{0.400pt}}
\put(494,370.67){\rule{0.723pt}{0.400pt}}
\multiput(494.00,371.17)(1.500,-1.000){2}{\rule{0.361pt}{0.400pt}}
\put(476.0,372.0){\rule[-0.200pt]{4.336pt}{0.400pt}}
\put(522,369.67){\rule{0.723pt}{0.400pt}}
\multiput(522.00,370.17)(1.500,-1.000){2}{\rule{0.361pt}{0.400pt}}
\put(497.0,371.0){\rule[-0.200pt]{6.022pt}{0.400pt}}
\put(559,368.67){\rule{0.723pt}{0.400pt}}
\multiput(559.00,369.17)(1.500,-1.000){2}{\rule{0.361pt}{0.400pt}}
\put(525.0,370.0){\rule[-0.200pt]{8.191pt}{0.400pt}}
\put(614,367.67){\rule{0.723pt}{0.400pt}}
\multiput(614.00,368.17)(1.500,-1.000){2}{\rule{0.361pt}{0.400pt}}
\put(562.0,369.0){\rule[-0.200pt]{12.527pt}{0.400pt}}
\put(704,366.67){\rule{0.723pt}{0.400pt}}
\multiput(704.00,367.17)(1.500,-1.000){2}{\rule{0.361pt}{0.400pt}}
\put(617.0,368.0){\rule[-0.200pt]{20.958pt}{0.400pt}}
\put(707.0,367.0){\rule[-0.200pt]{31.076pt}{0.400pt}}
\put(706,407){\makebox(0,0)[r]{$\Lambda^*$}}
\multiput(728,407)(20.756,0.000){4}{\usebox{\plotpoint}}
\put(794,407){\usebox{\plotpoint}}
\put(220,113){\usebox{\plotpoint}}
\put(220.00,113.00){\usebox{\plotpoint}}
\multiput(223,114)(19.690,6.563){0}{\usebox{\plotpoint}}
\multiput(226,115)(12.453,16.604){0}{\usebox{\plotpoint}}
\multiput(229,119)(12.453,16.604){0}{\usebox{\plotpoint}}
\put(234.28,126.80){\usebox{\plotpoint}}
\multiput(235,128)(9.282,18.564){0}{\usebox{\plotpoint}}
\multiput(238,134)(10.298,18.021){0}{\usebox{\plotpoint}}
\put(243.81,145.21){\usebox{\plotpoint}}
\multiput(245,148)(7.288,19.434){0}{\usebox{\plotpoint}}
\multiput(248,156)(7.288,19.434){0}{\usebox{\plotpoint}}
\put(251.22,164.60){\usebox{\plotpoint}}
\multiput(254,172)(7.288,19.434){0}{\usebox{\plotpoint}}
\put(258.51,184.03){\usebox{\plotpoint}}
\multiput(260,188)(7.288,19.434){0}{\usebox{\plotpoint}}
\put(265.80,203.46){\usebox{\plotpoint}}
\multiput(266,204)(7.288,19.434){0}{\usebox{\plotpoint}}
\multiput(269,212)(8.176,19.077){0}{\usebox{\plotpoint}}
\put(273.58,222.70){\usebox{\plotpoint}}
\multiput(275,226)(10.298,18.021){0}{\usebox{\plotpoint}}
\multiput(279,233)(9.282,18.564){0}{\usebox{\plotpoint}}
\put(283.07,241.14){\usebox{\plotpoint}}
\multiput(285,245)(9.282,18.564){0}{\usebox{\plotpoint}}
\multiput(288,251)(9.282,18.564){0}{\usebox{\plotpoint}}
\put(292.56,259.59){\usebox{\plotpoint}}
\multiput(294,262)(10.679,17.798){0}{\usebox{\plotpoint}}
\multiput(297,267)(10.679,17.798){0}{\usebox{\plotpoint}}
\multiput(300,272)(12.453,16.604){0}{\usebox{\plotpoint}}
\put(303.77,277.03){\usebox{\plotpoint}}
\multiput(306,280)(12.453,16.604){0}{\usebox{\plotpoint}}
\multiput(309,284)(12.453,16.604){0}{\usebox{\plotpoint}}
\multiput(312,288)(14.676,14.676){0}{\usebox{\plotpoint}}
\put(316.98,292.98){\usebox{\plotpoint}}
\multiput(319,295)(14.676,14.676){0}{\usebox{\plotpoint}}
\multiput(322,298)(14.676,14.676){0}{\usebox{\plotpoint}}
\multiput(325,301)(17.270,11.513){0}{\usebox{\plotpoint}}
\multiput(328,303)(14.676,14.676){0}{\usebox{\plotpoint}}
\put(332.30,306.87){\usebox{\plotpoint}}
\multiput(334,308)(14.676,14.676){0}{\usebox{\plotpoint}}
\multiput(337,311)(17.270,11.513){0}{\usebox{\plotpoint}}
\multiput(340,313)(17.270,11.513){0}{\usebox{\plotpoint}}
\multiput(343,315)(17.270,11.513){0}{\usebox{\plotpoint}}
\multiput(346,317)(17.270,11.513){0}{\usebox{\plotpoint}}
\put(349.04,319.03){\usebox{\plotpoint}}
\multiput(352,321)(20.136,5.034){0}{\usebox{\plotpoint}}
\multiput(356,322)(17.270,11.513){0}{\usebox{\plotpoint}}
\multiput(359,324)(19.690,6.563){0}{\usebox{\plotpoint}}
\multiput(362,325)(17.270,11.513){0}{\usebox{\plotpoint}}
\put(367.56,327.85){\usebox{\plotpoint}}
\multiput(368,328)(19.690,6.563){0}{\usebox{\plotpoint}}
\multiput(371,329)(17.270,11.513){0}{\usebox{\plotpoint}}
\multiput(374,331)(19.690,6.563){0}{\usebox{\plotpoint}}
\multiput(377,332)(19.690,6.563){0}{\usebox{\plotpoint}}
\multiput(380,333)(19.690,6.563){0}{\usebox{\plotpoint}}
\multiput(383,334)(19.690,6.563){0}{\usebox{\plotpoint}}
\put(386.83,335.28){\usebox{\plotpoint}}
\multiput(389,336)(19.690,6.563){0}{\usebox{\plotpoint}}
\multiput(392,337)(20.136,5.034){0}{\usebox{\plotpoint}}
\multiput(396,338)(19.690,6.563){0}{\usebox{\plotpoint}}
\multiput(399,339)(20.756,0.000){0}{\usebox{\plotpoint}}
\multiput(402,339)(19.690,6.563){0}{\usebox{\plotpoint}}
\put(406.77,340.59){\usebox{\plotpoint}}
\multiput(408,341)(19.690,6.563){0}{\usebox{\plotpoint}}
\multiput(411,342)(20.756,0.000){0}{\usebox{\plotpoint}}
\multiput(414,342)(19.690,6.563){0}{\usebox{\plotpoint}}
\multiput(417,343)(19.690,6.563){0}{\usebox{\plotpoint}}
\multiput(420,344)(20.756,0.000){0}{\usebox{\plotpoint}}
\multiput(423,344)(19.690,6.563){0}{\usebox{\plotpoint}}
\put(426.81,345.00){\usebox{\plotpoint}}
\multiput(429,345)(20.136,5.034){0}{\usebox{\plotpoint}}
\multiput(433,346)(20.756,0.000){0}{\usebox{\plotpoint}}
\multiput(436,346)(19.690,6.563){0}{\usebox{\plotpoint}}
\multiput(439,347)(20.756,0.000){0}{\usebox{\plotpoint}}
\multiput(442,347)(19.690,6.563){0}{\usebox{\plotpoint}}
\put(447.11,348.00){\usebox{\plotpoint}}
\multiput(448,348)(19.690,6.563){0}{\usebox{\plotpoint}}
\multiput(451,349)(20.756,0.000){0}{\usebox{\plotpoint}}
\multiput(454,349)(19.690,6.563){0}{\usebox{\plotpoint}}
\multiput(457,350)(20.756,0.000){0}{\usebox{\plotpoint}}
\multiput(460,350)(20.756,0.000){0}{\usebox{\plotpoint}}
\multiput(463,350)(19.690,6.563){0}{\usebox{\plotpoint}}
\put(467.38,351.00){\usebox{\plotpoint}}
\multiput(469,351)(20.756,0.000){0}{\usebox{\plotpoint}}
\multiput(473,351)(19.690,6.563){0}{\usebox{\plotpoint}}
\multiput(476,352)(20.756,0.000){0}{\usebox{\plotpoint}}
\multiput(479,352)(20.756,0.000){0}{\usebox{\plotpoint}}
\multiput(482,352)(19.690,6.563){0}{\usebox{\plotpoint}}
\put(487.81,353.00){\usebox{\plotpoint}}
\multiput(488,353)(20.756,0.000){0}{\usebox{\plotpoint}}
\multiput(491,353)(20.756,0.000){0}{\usebox{\plotpoint}}
\multiput(494,353)(19.690,6.563){0}{\usebox{\plotpoint}}
\multiput(497,354)(20.756,0.000){0}{\usebox{\plotpoint}}
\multiput(500,354)(20.756,0.000){0}{\usebox{\plotpoint}}
\multiput(503,354)(20.756,0.000){0}{\usebox{\plotpoint}}
\put(508.34,354.58){\usebox{\plotpoint}}
\multiput(510,355)(20.756,0.000){0}{\usebox{\plotpoint}}
\multiput(513,355)(20.756,0.000){0}{\usebox{\plotpoint}}
\multiput(516,355)(20.756,0.000){0}{\usebox{\plotpoint}}
\multiput(519,355)(20.756,0.000){0}{\usebox{\plotpoint}}
\multiput(522,355)(19.690,6.563){0}{\usebox{\plotpoint}}
\multiput(525,356)(20.756,0.000){0}{\usebox{\plotpoint}}
\put(528.88,356.00){\usebox{\plotpoint}}
\multiput(531,356)(20.756,0.000){0}{\usebox{\plotpoint}}
\multiput(534,356)(20.756,0.000){0}{\usebox{\plotpoint}}
\multiput(537,356)(20.756,0.000){0}{\usebox{\plotpoint}}
\multiput(540,356)(19.690,6.563){0}{\usebox{\plotpoint}}
\multiput(543,357)(20.756,0.000){0}{\usebox{\plotpoint}}
\put(549.47,357.00){\usebox{\plotpoint}}
\multiput(550,357)(20.756,0.000){0}{\usebox{\plotpoint}}
\multiput(553,357)(20.756,0.000){0}{\usebox{\plotpoint}}
\multiput(556,357)(20.756,0.000){0}{\usebox{\plotpoint}}
\multiput(559,357)(19.690,6.563){0}{\usebox{\plotpoint}}
\multiput(562,358)(20.756,0.000){0}{\usebox{\plotpoint}}
\multiput(565,358)(20.756,0.000){0}{\usebox{\plotpoint}}
\put(570.06,358.00){\usebox{\plotpoint}}
\multiput(571,358)(20.756,0.000){0}{\usebox{\plotpoint}}
\multiput(574,358)(20.756,0.000){0}{\usebox{\plotpoint}}
\multiput(577,358)(20.756,0.000){0}{\usebox{\plotpoint}}
\multiput(580,358)(20.756,0.000){0}{\usebox{\plotpoint}}
\multiput(583,358)(20.136,5.034){0}{\usebox{\plotpoint}}
\multiput(587,359)(20.756,0.000){0}{\usebox{\plotpoint}}
\put(590.70,359.00){\usebox{\plotpoint}}
\multiput(593,359)(20.756,0.000){0}{\usebox{\plotpoint}}
\multiput(596,359)(20.756,0.000){0}{\usebox{\plotpoint}}
\multiput(599,359)(20.756,0.000){0}{\usebox{\plotpoint}}
\multiput(602,359)(20.756,0.000){0}{\usebox{\plotpoint}}
\multiput(605,359)(20.756,0.000){0}{\usebox{\plotpoint}}
\multiput(608,359)(20.756,0.000){0}{\usebox{\plotpoint}}
\put(611.45,359.00){\usebox{\plotpoint}}
\multiput(614,359)(19.690,6.563){0}{\usebox{\plotpoint}}
\multiput(617,360)(20.756,0.000){0}{\usebox{\plotpoint}}
\multiput(620,360)(20.756,0.000){0}{\usebox{\plotpoint}}
\multiput(623,360)(20.756,0.000){0}{\usebox{\plotpoint}}
\multiput(627,360)(20.756,0.000){0}{\usebox{\plotpoint}}
\put(632.04,360.00){\usebox{\plotpoint}}
\multiput(633,360)(20.756,0.000){0}{\usebox{\plotpoint}}
\multiput(636,360)(20.756,0.000){0}{\usebox{\plotpoint}}
\multiput(639,360)(20.756,0.000){0}{\usebox{\plotpoint}}
\multiput(642,360)(20.756,0.000){0}{\usebox{\plotpoint}}
\multiput(645,360)(20.756,0.000){0}{\usebox{\plotpoint}}
\multiput(648,360)(20.756,0.000){0}{\usebox{\plotpoint}}
\put(652.71,360.57){\usebox{\plotpoint}}
\multiput(654,361)(20.756,0.000){0}{\usebox{\plotpoint}}
\multiput(657,361)(20.756,0.000){0}{\usebox{\plotpoint}}
\multiput(660,361)(20.756,0.000){0}{\usebox{\plotpoint}}
\multiput(664,361)(20.756,0.000){0}{\usebox{\plotpoint}}
\multiput(667,361)(20.756,0.000){0}{\usebox{\plotpoint}}
\multiput(670,361)(20.756,0.000){0}{\usebox{\plotpoint}}
\put(673.39,361.00){\usebox{\plotpoint}}
\multiput(676,361)(20.756,0.000){0}{\usebox{\plotpoint}}
\multiput(679,361)(20.756,0.000){0}{\usebox{\plotpoint}}
\multiput(682,361)(20.756,0.000){0}{\usebox{\plotpoint}}
\multiput(685,361)(20.756,0.000){0}{\usebox{\plotpoint}}
\multiput(688,361)(20.756,0.000){0}{\usebox{\plotpoint}}
\multiput(691,361)(20.756,0.000){0}{\usebox{\plotpoint}}
\put(694.15,361.00){\usebox{\plotpoint}}
\multiput(697,361)(20.756,0.000){0}{\usebox{\plotpoint}}
\multiput(700,361)(20.756,0.000){0}{\usebox{\plotpoint}}
\multiput(704,361)(19.690,6.563){0}{\usebox{\plotpoint}}
\multiput(707,362)(20.756,0.000){0}{\usebox{\plotpoint}}
\multiput(710,362)(20.756,0.000){0}{\usebox{\plotpoint}}
\put(714.74,362.00){\usebox{\plotpoint}}
\multiput(716,362)(20.756,0.000){0}{\usebox{\plotpoint}}
\multiput(719,362)(20.756,0.000){0}{\usebox{\plotpoint}}
\multiput(722,362)(20.756,0.000){0}{\usebox{\plotpoint}}
\multiput(725,362)(20.756,0.000){0}{\usebox{\plotpoint}}
\multiput(728,362)(20.756,0.000){0}{\usebox{\plotpoint}}
\multiput(731,362)(20.756,0.000){0}{\usebox{\plotpoint}}
\put(735.50,362.00){\usebox{\plotpoint}}
\multiput(737,362)(20.756,0.000){0}{\usebox{\plotpoint}}
\multiput(741,362)(20.756,0.000){0}{\usebox{\plotpoint}}
\multiput(744,362)(20.756,0.000){0}{\usebox{\plotpoint}}
\multiput(747,362)(20.756,0.000){0}{\usebox{\plotpoint}}
\multiput(750,362)(20.756,0.000){0}{\usebox{\plotpoint}}
\multiput(753,362)(20.756,0.000){0}{\usebox{\plotpoint}}
\put(756.25,362.00){\usebox{\plotpoint}}
\multiput(759,362)(20.756,0.000){0}{\usebox{\plotpoint}}
\multiput(762,362)(20.756,0.000){0}{\usebox{\plotpoint}}
\multiput(765,362)(20.756,0.000){0}{\usebox{\plotpoint}}
\multiput(768,362)(20.756,0.000){0}{\usebox{\plotpoint}}
\multiput(771,362)(20.756,0.000){0}{\usebox{\plotpoint}}
\multiput(774,362)(20.756,0.000){0}{\usebox{\plotpoint}}
\put(777.01,362.00){\usebox{\plotpoint}}
\multiput(781,362)(19.690,6.563){0}{\usebox{\plotpoint}}
\multiput(784,363)(20.756,0.000){0}{\usebox{\plotpoint}}
\multiput(787,363)(20.756,0.000){0}{\usebox{\plotpoint}}
\multiput(790,363)(20.756,0.000){0}{\usebox{\plotpoint}}
\multiput(793,363)(20.756,0.000){0}{\usebox{\plotpoint}}
\put(797.60,363.00){\usebox{\plotpoint}}
\multiput(799,363)(20.756,0.000){0}{\usebox{\plotpoint}}
\multiput(802,363)(20.756,0.000){0}{\usebox{\plotpoint}}
\multiput(805,363)(20.756,0.000){0}{\usebox{\plotpoint}}
\multiput(808,363)(20.756,0.000){0}{\usebox{\plotpoint}}
\multiput(811,363)(20.756,0.000){0}{\usebox{\plotpoint}}
\multiput(814,363)(20.756,0.000){0}{\usebox{\plotpoint}}
\put(818.36,363.00){\usebox{\plotpoint}}
\multiput(821,363)(20.756,0.000){0}{\usebox{\plotpoint}}
\multiput(824,363)(20.756,0.000){0}{\usebox{\plotpoint}}
\multiput(827,363)(20.756,0.000){0}{\usebox{\plotpoint}}
\multiput(830,363)(20.756,0.000){0}{\usebox{\plotpoint}}
\multiput(833,363)(20.756,0.000){0}{\usebox{\plotpoint}}
\put(836,363){\usebox{\plotpoint}}
\end{picture}

%% file: fermion.bbl
\begin{references}
\bibitem{ex}
As  example for $N=2$ the state $|0\rangle\otimes|1\rangle$ has no physical meaning
in that it amounts to the unphysical superposition $|\Psi\rangle^++|\Psi\rangle^-,\,(
|\Psi^\pm\rangle:=1/2\, ( |0\rangle\otimes|1\rangle\pm |1\rangle\otimes|0\rangle)).$
\bibitem{bose}
For Bosons (no constraints over  $N$)
the corresponding Fock space is the direct sum of
the $\pmatrix{N+L-1 \cr L-1}$-dimensional  totally symmetric sectors of
$h_L^{\otimes\,N}.$
This  direct sum is
isomorphic
to a tensor product:  $h_\infty^{\otimes\,L},$
where  $h_\infty$ denotes the  harmonic oscillator state-space.
\bibitem{not}
Notice that quantity  (\ref{aim}) is different
from thermal entanglement as discussed in Ref \cite{thermal}.
The latter being defined as the mixed-state entanglement of the
thermal Gibb's state
\bibitem{eq_heis}
From Eq (\ref{hub_gs}) it follows that for $U\mapsto\infty$ the dimer
ground state  mapsto onto the spin $1/2$  singlet
with (canonical) energy  
$\epsilon_0\sim -4\,t^2/U.$
\end{references}
